%Paper: hep-ph/9206233
%From: bennett@sojin.llnl.gov (Dave Bennett)
%Date: Thu, 18 Jun 92 16:32:58 PDT

\input phyzzx
\def \header#1{\goodbreak\bigskip\centerline{\bf #1}\medskip\nobreak}
\def\s {\scriptscriptstyle}
\def\mathrelfun#1#2{\lower3.6pt\vbox{\baselineskip0pt\lineskip.9pt
  \ialign{$\mathsurround=0pt#1\hfil##\hfil$\crcr#2\crcr\sim\crcr}}}

\def\ln {{\rm ln}}
\def\hatbfn  {{\hat{\bf n}}}
\def\rmH  {{\rm H}}
\def\rmi  {{\rm i}}
\def\FNAL{\address{Fermi National Accelerator Laboratory\break
      P. O. Box 500, Batavia, Illinois, 60510}}
\def\IGPP{\address{Institute for Geophysics and Planetary Physics,
      Lawrence Livermore National Laboratory, Livermore, CA 94550}}

\PHYSREV
\singlespace
%%\doublespace

\pubnum{hep-ph/9206233}
\pubtype{}
\titlepage
\singlespace
%%\doublespace
\title{THE IMPLICATIONS OF THE COBE-DMR RESULTS FOR COSMIC STRINGS}
\author{David P. Bennett}
\IGPP
\andauthor{Albert Stebbins}
\FNAL
\andauthor{Fran\c cois R. Bouchet}
\address{Institut d'Astrophysique de Paris, CNRS,
         98 bis Blvd Arago, 75014 Paris}

\abstract
\singlespace
%%\doublespace
We compare the anisotropies in the cosmic microwave background radiation
measured by the COBE experiment to the predictions of cosmic strings.
We use an analytic model for the $\Delta T/T$ power spectrum that is
based on our previous numerical simulations to show that the COBE
results imply a value for the string mass per unit length, $\mu$ under the
assumption that cosmic strings are the source of the measured anisotropy.
We find $G\mu = 1.5\pm 0.5 \times 10^{-6}$ which is consistent with
the value of $\mu$ thought to be required for cosmic strings to seed
galaxy formation.

\submit{The Astrophysical Journal Letters}
\endpage

\unnumberedchapters
\chapter{}

%%% Figure captions
%%%%%%%%%%%%%%%%%%%
\FIG\figone{The spectral index $n = {d\ln C_l\over dl}+3$, the Gaussian weight
function due to the COBE-DMR beam smearing and the best fit value of $G\mu$
are plotted as a function of $l$ assuming the COBE best fit power spectrum for
$n=1.4$.}

%%%  Main text
%%%%%%%%%%%%%%

\header{I. Introduction}

	Recently the COBE (COsmic Background Explorer) collaboration has
announced brightness fluctuations in the sky at centimeter wavelengths (Smoot
\etal, 1992) as measured by the DMR experiment.  These fluctuations do not have
the characteristics expected of emission from foreground objects at low
redshifts and are believed to represent intrinsic temperature fluctuations in
the cosmic microwave background radiation (MBR).  Henceforth we will assume
that the brightness fluctuations are intrinsic temperature fluctuations (i.e.
MBR anisotropies).  The result is very exciting as these intrinsic fluctuations
are perhaps the only available probe of cosmological structures on the scale of
thousands of megaparsecs.  Determining the nature of structures on these large
scales can provide important hints as to the primordial fluctuations on smaller
scales which grew to form objects from stars and planets to superclusters of
galaxies.  The large scale fluctuations also give important hints as to the
very early history of our universe.

	At present only details of the spectrum ($\equiv$ angular correlation
function) of temperature perturbations have been announced.  This is not enough
information to determine the detailed nature of the production of the
temperature perturbations.  In the usual picture of primordial adiabatic
perturbations, the temperature fluctuations arise from the "gravitational
redshift" effects associated with primordial potential hills and valleys.  If
this is the case then the COBE results indicate that the initial spectrum of
perturbations is approximately of the Harrison-Zel'dovich form (Harrison, 1970,
and Zel'dovich, 1972).  Another possibility is that the temperature
fluctuations arose from the time varying gravitational field associated with
the motion of seeds, such as cosmic strings, global monopoles, or cosmic
textures (Kibble, 1976, Vilenkin, 1980, Zel'dovich, 1980, Turok, 1989, Bennett
and Rhie, 1990).  The interpretation of the COBE results in these two types of
models are very different, although the method for calculating the anisotropies
is essentially the same (Sachs and Wolfe, 1967).  For primordial perturbations
COBE is seeing density fluctuation on the surface of last scattering which are
correlated on scales much greater than the ``apparent'' causal horizon at that
time.  In the case of seeded perturbations, the temperature fluctuations on
large scales are induced at times long after last scattering, and no
``acausal'' correlations are required. In spite of this difference the two
types of theories can produce very similar temperature fluctuations.  For
example, all of the seed models mentioned above lead to a final spectrum of
density and temperature fluctuations which are at least approximately of the
Harrison-Zel'dovich form on the scales observed by COBE (Albrecht and Stebbins,
1992a and 1992b, Park, Spergel and Turok, 1991, and Bennett, Rhie and Weinberg,
1992).

	Thus, the COBE results should be viewed as encouraging from the point of
view of proponents of seeded perturbations.  However COBE not only measures the
shape of the spectrum of perturbations, but also the amplitude.  For the theory
to be acceptable the amplitude that is measured should be the same as is needed
to form the observed structures in the universe.  While our ignorance of galaxy
formation may leave us a little fuzzy as to the exact amplitude that is
required, there is still only a range of amplitudes which might be considered
acceptable.  In this {\it Letter} we will consider the cosmic string scenario,
and find that our best estimate of the amplitude of perturbations indicated by
COBE is within the acceptable range.  Textures and global monopoles are
considered elsewhere (Turok and Spergel, 1990, Bennett and Rhie, 1992).
Another issue which we do not address in this paper is whether the pattern of
anisotropy is consistent with the predictions of cosmic strings.  The COBE
collaboration has not yet released enough information to determine this, and if
they had, it seems likely that the signal to noise would be too low to
distinguish the non-Gaussian character of cosmic string induced anisotropies.

\header{2. Constraints of String Model from First Year of COBE DMR}

	What does COBE say about strings, assuming they are the cause of the
anisotropy?  The amplitude of the perturbations produced by cosmic strings is
proportional to the mass per unit length $\mu$ of the strings.  So from the
COBE results we should be able to normalize this parameter.  As with other
models, we must rely on the existence of a preponderance of non-baryonic dark
matter at the time of last scattering in order that the small scale
perturbations not be washed out by sound waves.  The nature of the dark matter
is an important parameter of the string model.  It may be hot dark matter (HDM)
such as a light massive neutrino, or cold dark matter (CDM) such as an axion.
The shape of the COBE fluctuation spectrum will not tell us anything about
which type of dark matter there is, because on the scales that COBE probes,
both HDM and CDM act essentially the same.  However the value of $\mu$ we
obtain may give us some clue as to the nature of the dark matter.

	The best available estimates of MBR anisotropy from strings are those
of Bouchet, Bennett, and Stebbins (1988, hereafter BBS).  Several groups are
working to improve on these calculations but no results are yet available.  The
problem with the BBS calculation is they have used a formalism appropriate for
the small angle and for strings in Minkowski space (Stebbins 1988).  Clearly
neither is completely appropriate for the COBE experiment.  However the results
are not completely inappropriate either.  For photons coming much closer than a
horizon distance from a piece of string BBS should be accurate. We will make
some adjustments of the results of BBS for the deviations from the small angle
approximation, but corrections for expansion are much more complicated and will
not be attempted.  Thus, the reader should keep in mind that there remain
significant systematic errors in the results we will present.

	The results in BBS for the power spectrum of temperature fluctuation is
given by the fitting BBS equation (4) which may be rewritten:
$$\int_0^\lambda d\lambda\int_{a_{\rm i}}^{2a_{\rm i}}da\,
{dC(0)\over d\lambda\,da}=A^2\,F\left({\lambda\over\Theta_{\rmH_\rmi}}\right)
\quad A=6{G\mu\over c^2} \quad
%%% \qquad A=6{G\mu\over c^2} \qquad
       F(x)=\left({x^{1.7}\over (0.6)^{1.7}+x^{1.7}}\right)^{0.7}
,\eqn\BBSeq$$
where $C(0)$ gives the mean-square ${\Delta T\over T}$, $a$ is the scale factor
when these temperature fluctuations are produced, $\Theta_{\rm Hi}$ is the
tangent of the angle subtended,
perpendicular to the line-of-sight, by the horizon
when $a=a_\rmi$, and $\lambda$ refers to the angular wavelength in radians of a
Fourier decomposition of the temperature pattern in the small angle
approximation.  We may approximate equation \BBSeq\ by
$${d C(0)\over d\ln\lambda\,d\ln a}=
                   {A^2\over\ln2}{\lambda\over\Theta_{\rmH_\rmi}}
                    F'\left({\lambda\over\Theta_{\rmH_\rmi}}\right)
\eqn\eqtwo$$
where $C(\theta)$ is the angular correlation function of the temperature
anisotropy.

If we work in units where $a_{\rm now} = 1$, then we may
use $\Theta_{\rmH_\rmi} = \sqrt{a}/(1-\sqrt{a})$  and we find
$$\eqalign{
{d C(0)\over d\ln\lambda}
=&{A^2\over\ln2}\int_{a_{\rm ls}}^1 {\lambda\over\Theta_{\rmH_\rmi}}
          F'\left({\lambda\over\Theta_{\rmH_\rmi}}\right)\,{d a\over a} \cr
=& {2A^2\over\ln2}
\left[(1-\sqrt{a_{\rm ls}})\,F\left({\lambda\over\Theta_{\rmH_{\rm ls}}}\right)
-\int_{\sqrt{a_{\rm ls}}}^1
                      F\left({\lambda\over\Theta_{\rmH_\rmi}}\right)\,d\sqrt{a}
        \right]    },\eqn\eqthree $$
where ``ls'' refers to the last-scattering surface.  When $\Theta_{\rmH_{\rm
ls}}\ll\lambda\ll1$ the $[\cdots]$ in eq. \eqthree\ goes to unity and we can
see that the temperature power spectrum is ``scale-invariant''. Physically,
this limit corresponds to the case where all the contributions to $\Delta T/T$
come from strings at $\sqrt{a_{\rm ls}} \ll \sqrt{a} \ll 1$. The effects due to
non-zero $a_{\rm ls}$ are negligible on the COBE scales, however deviations
from scale invariance as one goes beyond $\lambda\ll1$ are not. An important
deviation from scale invariance which is due to the contribution of low
redshift strings to small angle anisotropies is properly included in Eq.
\eqthree\ .

	A Fourier decomposition on small scales corresponds to a decomposition
into spherical harmonics on the sphere of the sky:
$${\Delta T\over T}(\hatbfn)
        =\sum_{l=0}^\infty\sum_{m=-l}^l a_{\s(l,m)}Y_{\s(l,m)}(\hatbfn)
\eqn\eqfour$$
where since $\nabla^2 Y_{\s(l,m)}=-l(l+1)Y_{\s(l,m)}$ we see that $\lambda$
becomes $2\pi/\sqrt{l(l+1)}$ and only a discrete set of wavelengths are
allowed.  The analog of the power spectra is
$$C_l=\sum_{m=-l}^l |a_{\s(l,m)}|^2
\eqn\eqfive $$
which gives the mean square temperature fluctuation is
$$C(0)=\sum_{l=0}^\infty {2l+1\over4\pi}C_l
.\eqn\eqsix $$
For large $l$ we make take the sum to an integral obtaining
$${d C(0)\over d\ln l}={d C(0)\over d\ln\lambda}={l(l+1)\over2\pi}C_l
 \qquad l\gg1
\eqn\eqseven $$
which may be compared to Eq. \eqthree.

Smoot \etal (1992) have fit their results to ``power law" correlation functions
of the form
$$C_l=(Q_{rms-PS})^2 {4\pi\over 5}
         {\Gamma(l+{n-1\over2})\,\Gamma({9-n\over2})\over
          \Gamma(l+{5-n\over2})\,\Gamma({3+n\over2})} \ .
\eqn\eqCl$$
When they take into account the cosmic variance, and do not include the
measured \nextline
quadrupole in their fit, they obtain best fit values of
$n = 1.15 {+0.45 \atop -0.65}$ and $Q_{rms-PS} = 5.96\pm 1.68\times 10^{-6}$.
When they include the quadrupole, they obtain
$n = 1.5 $ and $Q_{rms-PS} = 5.1 \times 10^{-6}$.

In the extreme small angle approximation, cosmic strings predict a spectral
index of $n=1$, but for the range of $l$ that COBE is sensitive to
and for which we expect our calculations ($5 \lsim l \lsim 20$) to be valid,
we find a spectral index closer to $n=1.4$ (see Fig. \figone).
The reason that we find $n > 1$ at large scales is that much of the power at
a given scale does not get generated until that scale is considerably
smaller than the horizon, so power is missing at large scales because it
has yet to be generated. It is interesting to note that our string induced
power spectrum is closer to the best fit to COBE (when the
quadrupole is included in the fit) than inflationary prediction of $n=1$.

In order to compare our predicted amplitude with the COBE results, let us
consider the COBE fits for fixed $n$:
$Q_{rms-PS} = 6.11\pm 1.68\times 10^{-6}$ for $n=1$ (Smoot \etal, 1992), and
$Q_{rms-PS} = 4.75\pm 1.30\times 10^{-6}$ for $n=1.4$ (G. Smoot, private
communication). For the $n=1$ case and the extreme small angle approximation,
we can neglect the integral in eq. \eqthree, and then use eqs. \BBSeq,
\eqseven, and \eqCl\ to find
$$ G\mu = {1\over 6} \sqrt{6\ln 2\over 5} Q_{rms-PS} = 0.152\,Q_{rms-PS}
        = 9.3\pm 2.6\times 10^{-7} \ .
\eqn\Gmuesaa $$
A more accurate value can be obtained by using the $n=1.4$ fit values to
compare with eq. \eqthree\ evaluated numerically. The results of this
calculation are displayed in Fig. \figone\ which shows the
results of the integral in eq. \eqthree, and the resulting values
of $G\mu$. Our best fit value of $G\mu \simeq 1.49\times 10^{-6}$
is shown as a function of $l$
because the $n=1.4$ power law is not a perfect fit to our results.
Also plotted in Fig. \figone\ is the Gaussian weight function
$W^2(l) = e^{-l(l+1)/17.8^2}$ which describes the smoothing due to the
COBE-DMR $7^\circ$ beam.

Now let us attempt to estimate the errors in our calculation. These
errors come from two main sources: the small angle approximation,
and the Minkowski space approximation. We expect that the errors due
to the small angle approximation are small if we restrict ourselves
to moderately large $l$ values (say $l > 5$). The errors due to the
Minkowski space approximation are not so easy to quantify, but one
obvious symptom of this approximation is the power spectrum given in
eq. \BBSeq\ which extends outside the horizon. In a complete treatment
of this problem, one would expect that the power spectrum might get
cut off at around $\lambda =\Theta_{\rmH_\rmi} $ due to the required
compensation of the string perturbations by the matter fields. Inserting
this cutoff in our calculation gives $G\mu \simeq 1.85\times 10^{-6}$.
Although one might expect that adding the effect of compensation would always
tend to decrease $\Delta T/T$ and therefore increase our estimate of
$G\mu$, there are cases in which adding in the compensation actually
serves to increase $\Delta T/T$ (Stebbins and Veeraraghavan, 1992).
Therefore, we will take
$(1.85-1.49)\times10^{-6}= 0.36\times 10^{-6}$ to be our (symmetric) 1-$\sigma$
error bars. Adding these in quadrature  with COBE's experimental error bars
yields $G\mu = 1.49\pm 0.52\times10^{-6}$ which is our prediction.

How does this prediction for $G\mu$ compare with smaller angular scale
experiments? The answer to this question is muddied somewhat by the
possibility that the universe underwent reionization, but some conclusions
are still possible. A detailed comparison with the
OVRO experiment (Readhead, \etal, 1989) indicates a rather weak limit,
$G\mu \lsim 4\times 10^{-6}$ due in part to the non-Gaussian character of
the anisotropies on arc minute scales (Bennett, Bouchet, and Stebbins, 1992).
In a reionized universe, the OVRO limit would be much weaker. More intriguing
is a comparison with the MAX experiment which has likely detected anisotropy
on a $1^\circ$ scale (Devlin, \etal., 1992). (They have a strong signal which
is consistent with CMB anisotropy, but they do not make a definitive claim
that this is what they have detected.) A preliminary analysis of their
data indicates that it is consistent with our value for $G\mu$ in models
both with and without early reionization.

	Another observational constraint on $\mu$ comes from the gravitational
waves they produce, which can cause ``jitter'' in the timing of rapidly
rotating pulsars. Our predicted value of $\mu$ is more than an order of
magnitude below the upper limits set by Bennett and Bouchet (1991). The
Bennett-Bouchet limit has the advantage that it has very little dependence
on poorly understood details of cosmic string evolution, but
Caldwell and Allen (1991) have shown that a considerably more stringent limit
is possible when one assumes a specific model for cosmic string evolution.
The most stringent limits on $\mu$ found by Caldwell and Allen should be
regarded with some skepticism, however, since they rely on an unpublished
analysis of the pulsar timing data (Ryba, 1991). (Previous unpublished
analyses of the pulsar timing data (Taylor, 1989) have resulted in
``limits" that have subsequently been revised upward by an order of
magnitude (Stinebring, \etal, 1990).) In addition, Caldwell and Allen
have ignored the possibility that infinitely long cosmic strings might radiate
a significant amount energy directly into gravity waves as claimed by
Allen and Shellard (1992). If true, this would serve to weaken the Caldwell
and Allen bounds on $\mu$ by a factor of $\sim 2$. If we accept the unpublished
pulsar timing analysis and revise Allen and Caldwell's limits upward by
a factor of 2 to account for the radiation from long strings,
then we find that the Allen and Caldwell limits are
$G\mu \leq 6\times 10^{-7}$ for $h=1$ and $G\mu \leq 1.6\times 10^{-6}$ for
$h=0.75$. Thus, if $h=1$ and one accepts the unpublished analysis,
then our fit to the COBE data is almost inconsistent with the pulsar timing
data at the $2\,\sigma$ level. If we consider only the published pulsar
timing analysis or smaller values of $h$, then there is good agreement
between our fit to COBE and the pulsar timing data.

\header{3. Conclusions}

	In this {\it Letter} we have shown that the recent COBE results are
quite consistent with the idea that inhomogeneities in our universe were
induced by cosmic strings in an flat FRW cosmology, which is predominantly dark
matter. If this is so then we estimate that the mass per unit length of the
strings is $1.49\pm 0.52\times10^{-6}$.  This can be compared to the
value of $G\mu$ thought to be required in order to seed galaxy formation.
The most sophisticated calculation to date of the values of $G\mu$ required for
string seeded galaxy formation scenarios have been done by Albrecht
and Stebbins (1992a and 1992b). They estimate
$G\mu\approx 2.0\times 10^{-6}/b_8$ for the $h=1$ hot dark matter (HDM) model,
$G\mu\approx 4.0\times 10^{-6}/b_8$ for the $h=0.5$ HDM model,
$G\mu\approx 1.8\times 10^{-6}/b_8$ for the $h=1$ cold dark matter (CDM) model,
and $G\mu\approx 2.8\times 10^{-6}/b_8$ for the $h=0.5$ CDM model.
$h = H_0/(100 {\rm km/sec\,Mpc^{-1}})$ and $b_8$ is the bias factor which
gives the normalization of the density field: the RMS density fluctuation in
a sphere of radius $8\,h^{-1}\,$Mpc is $1/b_8$. Thus, values of $b_8$
between 1 and 3 seem to be consistent with the COBE-DMR results. This is
roughly the range of values that are considered to be plausible from
considerations of galaxy formation. In contrast, the standard CDM model
for galaxy formation seems to require much smaller values of $b_8$. The
COBE measurement implies $b_8 = 0.90\pm 0.25$ for $h=0.5$, $\Omega_b =0.03$;
$b_8 = 0.59\pm 0.16$ for $h=0.75$, $\Omega_b =0.03$; and
$b_8 = 1.03\pm 0.28$ for $h=0.5$, $\Omega_b =0.1$ for the standard CDM
model according to the calculations of Bond and Efstathiou (1987). Since
$1.5 \lsim b_8 \lsim 2.5$ is generally thought to be required for standard
CDM, the COBE data is only marginally consistent with this model. Thus,
if the calculations presented here and the calculations of Albrecht and
Stebbins (1992a and 1992b) are confirmed by more detailed work, we can
conclude that the COBE-DMR anisotropy measurements favor cosmic string
models over standard CDM.

\singlespace
\ack
The work of DPB was supported in part by the U.S. Department of Energy at the
Lawrence Livermore National Laboratory under contract No. W-7405-Eng-48.  AS
was supported in part by the DOE and the NASA through grant number NAGW-2381.

%%\endpage

\def\pp{\parshape 2 0truecm 16.25truecm 2truecm 14.25truecm}
\def\newrefout{\par \penalty-400 \vskip\chapterskip
   \spacecheck\referenceminspace \immediate\closeout\referencewrite
   \referenceopenfalse
   \line{\fourteenrm\hfil REFERENCES\hfil}\vskip\headskip
   }
\newrefout
\parskip=0pt
\pp\par
Albrecht, A., and Turok, N., 1989, Phys. Rev. {\bf D40}, 973.
\pp\par
Albrecht, A., and Stebbins, A., 1992a Phys. Rev. Lett. {\bf 68}, 2121.
\pp\par
Albrecht, A., and Stebbins, A., 1992b, Fermilab preprint.
\pp\par
Allen, B., and Shellard, E. P. S., 1992, Phys. Rev. {\bf D45}, 1898.
\pp\par
Bennett, D. P., and Bouchet, F. R., 1991, Phys. Rev. {\bf D43}, 2733.
\pp\par
Bennett, D. P., Bouchet, F. R., and Stebbins, A., 1992, in preparation.
\pp\par
Bennett, D. P., and Rhie, S. H., 1990, Phys. Rev. Lett. {\bf 65}, 1709.
\pp\par
Bennett, D. P., and Rhie, S. H., 1992, IGPP preprint.
\pp\par
Bennett, D. P., Rhie, S. H., and Weinberg, D. H., 1992, in preparation.
\pp\par
Bond, J. R., and Efstathiou, 1987, Mon. Not. R. Astron. Soc. {\bf 226}, 655.
\pp\par
Bouchet, F. R., Bennett, D. P., and Stebbins, A., 1988, Nature {\bf 335}, 410.
\pp\par
Devlin, M., \etal, 1992, submitted to Proc. Nat. Aca. Sci.
\pp\par
Caldwell, R. R., and Allen, B., 1991, preprint WISC-MILW-91-TH-14.
\pp\par
Harrison, E. R., 1970, Phys. Rev. {\bf D1}, 2726.
\pp\par
Kibble, T. W. B., 1976, J. Phys. {\bf A9}, 1387.
\pp\par
Park, C., Spergel, D. N., and Turok, N., 1990, Astrophys. J., {\bf 372}, L53.
\pp\par
Readhead, A. C., \etal., 1989, Astrophys. J. {\bf 346}, 566.
\pp\par
Ryba, M. F., 1991, unpublished.
\pp\par
Sachs, K., and Wolfe, A. M., 1967, Astrophys. J., {\bf 147}, 73.
\pp\par
Smoot, G., \etal, 1992, COBE preprint.
\pp\par
Stebbins, A., 1988, Ap. J., {\bf 327}, 584.
\pp\par
Stebbins, A. and Veeraraghavan, S., 1992, Ap. J. {\it Lett.} to appear.
\pp\par
Stinebring, D. R., Ryba, M. F., Taylor, J. H., and Romani, R. W., 1990,
 Phys. Rev. Lett. {\bf 65}, 285.
\pp\par
Taylor, J. H., 1989, unpublished.
\pp\par
Turok, N., 1989, Phys. Rev. Lett. {\bf 63}, 2625.
\pp\par
Turok, N., and Spergel, D. N., 1990, Phys. Rev. Lett. {\bf 64}, 2736.
\pp\par
Vilenkin, A., 1980, Phys. Rev. Lett. {\bf 46}, 1169, 1496(E).
\pp\par
Zel'dovich, Y. B., 1972, Mon. Not. R. Astron. Soc. {\bf 160}, 1P.
\pp\par
Zel'dovich, Y. B., 1980, Mon. Not. R. Astron. Soc. {\bf 192}, 663.

\figout
\end